\documentclass[notitlepage,twocolumn,superscriptaddress]{revtex4-1}

\usepackage{graphicx}
\usepackage{amsmath}
\usepackage[utf8]{inputenc}
\newcommand{\id}{\textrm{d}}
\usepackage{hyperref}

 \let\be=\beta  
\let\ve=\varepsilon   
 \let\la=\lambda \let\om=\omega

\let\De=\Delta  \let\La=\Lambda \let\Om=\Omega

\begin{document}

\title{Driving-induced stability with long-range effects}

\author{Urna Basu}
\affiliation{SISSA - International School for Advanced Studies,  Trieste, Italy}
\author{Pierre de Buyl}
\affiliation{Instituut voor Theoretische Fysica, KU Leuven, Belgium}
\author{Christian Maes}
\affiliation{Instituut voor Theoretische Fysica, KU Leuven, Belgium}
\author{Karel Netočný}
\affiliation{Institute of Physics, Academy of Sciences of the Czech Republic, Prague, Czech Republic}

\begin{abstract}
We give a sufficient condition under which an applied rotation on medium particles stabilizes a slow probe in the rotation center.
The symmetric part of the stiffness matrix thus gets a
  positive Lamb shift with respect to
  equilibrium. For illustration we take diffusive medium
  particles with a self-potential in the shape of a Mexican hat, high around the origin. There is a short-range attraction between the medium particles and the heavier probe, all immersed in an equilibrium thermal bath.
  For no or small rotation force on the medium particles, the origin is an unstable fixed point for the probe and
  the precise shape of the self-potential at large distances from the origin is irrelevant for
  the statistical force there.  Above a certain rotation threshold, while the medium particles are still repelled from the origin,
 the probe stabilizes there and more details of the medium-density at large distance start to matter.  The effect is robust around the quasi-static limit with rotation threshold only weakly depending on the temperature but the stabilization gets stronger at lower temperatures.
\end{abstract}

\maketitle

Stabilizing an otherwise unstable configuration or phase by external action is an important challenge for a range of applications but also for the physical understanding of spatio-temporal patterns induced by nonequilibrium effects. Many examples exist for dynamical systems where by using
feedback mechanisms one achieves the necessary control or steering.
Other examples such as the Kapitza (inverted) pendulum which is stabilized by a time-dependent external force do not require feedback~\cite{kapitza_1965a,krieger_kapitsa_arxiv_2015}.
A further step would be to eliminate the time-dependence and to use the steady nonequilibrium character of a medium to achieve such a
stabilization, possibly leading to robust time-independent control strategies.  The fact that the medium is quasi-stationary is relevant for the occurrence of stable structures in living matter~\cite{marchetti_soft_active_review_2013}, or also in collective Hamiltonian dynamics~\cite{barre_prl_2002}.

In the context of statistical forces, one aims at understanding the action of an ensemble of
particles on some collective coordinate or probe. Such a force can be derived consistently in
equilibrium statistical mechanics as the derivative of a free energy.
In the present letter we study the statistical force from a nonequilibrium medium on a slow
probe. While it can be viewed as an application of the formalism that has been introduced in refs.~\cite{basu_prl_2015,basu_njp_2015}, we concentrate here on 
driving-induced stabilization of a fixed point for the probe's dynamics. We consider a two-dimensional set-up with overdamped particles being driven by a solenoid flow and connected to a slow probe.  The stabilization of the probe at the rotation center is described by positive changes in the stiffness matrix.  In fact the symmetric part of the stiffness matrix is given in terms of a covariance between excess work functions. That excess work involves the nonequilibrium density globally and not only at the location of the probe, which signifies long range effects on the nature of the stabilization.

  We illustrate the theory with the example of driven particles confined by a Mexican-hat like potential. The origin is a fixed point for the probe but for attractive interaction with the medium is unstable in equilibrium; it acquires stability when increasing the rotational driving beyond a threshold value. It is important to note here that the medium's radial density-profile is almost not affected by the rotation.

Using numerical simulations, we also investigate more quantitative issues and how the phenomenon remains present beyond the (theoretical) quasi-static limit.  We also find that the phenomenon is robust with respect to changes in the driving (differential rotation) and we investigate the temperature-dependence of the effective spring constant.

We start by giving the general coupled dynamics of medium and probe that enables to ask for statistical forces and their corresponding stiffness matrix in the limit of a quasi-static probe.
We then state our main sufficient condition and result on the positivity of the (nonequilibrium) Lamb shift, i.e., on the stabilizing effect of the nonequilibrium driving. 
Because of long-range effects, linearization of the medium dynamics is not allowed in the nonequilibrium regime to reproduce the stiffness of the statistical force on the probe.
We present a specific illustration for a medium in a Mexican hat-like landscape, which also allows to explore the stabilization numerically beyond the quasi-static regime.

\section{Coupled dynamics, statistical force and stiffness}
We consider a two-dimensional system in which $N$ driven particles and a probe move in a thermal environment at temperature $T$, idealized here by using an overdamped Langevin dynamics. We refer to the driven particles at positions $y^i$ as the medium, which are mutually noninteracting and subject to a sufficiently confining potential $V(y)$. Each interacts with the probe via the potential $U_I(|x-y|)$ depending on the distance to the probe at position $x$. We write $U(x,y) = V(y) + U_I(|x-y|)$ for the total potential.  Furthermore, each of the medium particles is subject to a solenoidal driving force $F(y)$. The mobility for the medium particles is denoted by $\chi > 0$ and the damping coefficient for the probe is $\gamma>0$, so that the joint dynamics becomes, for $i=1,\ldots,N$,
\begin{eqnarray}
\dot y_t^i = \chi \left[ F(y_t^i) - \nabla_y U(x_t,y_t^i)\right] + \sqrt{2T\chi} \;\xi_t^i,\,
\nabla \cdot F = 0  \label{diff} \\
\gamma \dot{x}_t = - \sum_{i=1}^N\nabla_{x}  U_I(|x_t-y_t^i|) +\sqrt{2\gamma\,T}\; \xi_t \qquad \qquad  \qquad \label{probedyn}
\end{eqnarray}
all smoothly depending on the positions and under free boundary conditions at infinity. The $\xi^i_t,\xi_t$ are independent standard white noises.  Later for convenient simulation we add also a self-potential $V_p$ on the probe.\\
  We assume that the origin is a special point of symmetry, in the sense that when $x=0$ (probe at the origin) the force $F$ is always orthogonal to the force $\nabla_y U(x=0,y^i)$ on the medium particles. An example is provided by particles in a rotation symmetric self-potential $V(y^i)=V(|y^i|)$ which are driven by a rotational driving force $F$ having only an angular (and no radial) component around the origin.

The quasi-static regime for the probe is reached when the medium has a very small relaxation time compared to the probe, or $\gamma\chi \to \infty.$  (Below we also explore the joint dynamics when the time-scale separation between (fast) medium and (slow) probe is not infinite.) 
The main object of study is then the statistical force
\begin{eqnarray}\label{qsf}
f(x) &=& - \int \prod_{i=1}^N \left(\id y^i \rho_x(y^i)\right)\, \sum_{i=1}^N \nabla_x U(x,y^i)\nonumber\\
 &=& -N\, \int\, \id y \,\rho_x(y)\, \nabla_x U(x,y)\nonumber\\
 &=& -N\,\langle \nabla_x U_x\rangle^{x}
\end{eqnarray}
where we average over the stationary medium density $\rho_x(y)$ for a single driven particle.  We also write $U_x(y) =U(x,y)$ and $\langle\cdot\rangle^x$ is the expectation over $\rho_x$.  We always have the origin to be a fixed point in the
sense that the statistical force $f(x=0)=0$ vanishes there.  The statistical force has of course various components $f=(f_k)$ depending on the decomposition in orthogonal coordinates.\\

To investigate the stability of the probe near the origin, we introduce the stiffness matrix; see the beginning of the Appendix.
The stiffness at $x=0$ is defined by the matrix
\begin{equation}\label{sti}
M_{jk} = -\partial_j f_k(0) = \partial_j \langle \partial_k U_0 \rangle^0
\end{equation}
where we employ the notation
$\partial_j u_0 = (\partial u_x / \partial x_j)|_{x=0}$; analogously for $\partial_j \langle u_0 \rangle^0$ or
$\partial_j \partial_k u_0$.  A sufficient condition for local stability~\cite{khalil_nonlinear_systems} is the positivity of the stiffness matrix \eqref{sti}, which however only depends on its symmetric part.  The main subject of the paper is to understand how for the probe the origin stiffens under nonequilibrium.\\

Under equilibrium, for $F = 0$, the statistical force derives from the free energy ${\mathcal F}(x) = - TN\nabla_x\log {\mathcal{Z}}_x $, where the
partition function $\mathcal{Z}_x$ is, as ever,
\begin{equation*}
  {\mathcal Z}_x = \int \id y ~ \exp \left[ -\beta U(x,y) \right]
\end{equation*}
for $\beta = 1/T$.
The equilibrium stiffness ($F=0$) is
\begin{equation}\label{stiff-eq}
M_{jk}^\text{eq} = \be^{-1} \partial_j \partial_k \log {\cal Z}_0
\end{equation}
which is automatically symmetric (Maxwell relations), but can be negative in which case the probe is not stable at the origin.\\

When we are away from the quasi-static regime, we can still look at the total force on the probe in the joint (medium plus probe) steady ensemble.
We then consider the conditional expectation
\begin{equation}\label{condf}
g(X) = -\langle \nabla_x U(x,y) ~|~ x=X\rangle
\end{equation}
which in the quasi-static limit or in equilibrium coincides with \eqref{qsf}, $g(x) = f(x)$; not otherwise however. Note that {in nonequilibrium there is no reason for the force $f$ to be derived from the effective potential $V_{\text{eff}}(X)= - T \log \langle \delta(x-X) \rangle.$

\section{Nonequilibrium Lamb shift in the quasi-static limit}
The stiffness  \eqref{sti} at $x=0$ equals
\begin{equation}\label{int}
M_{jk} = \langle \partial_j \partial_k U_0 \rangle^0 +
\langle (\partial_j \log\rho_0)\, (\partial_k U_0) \rangle^0
\end{equation}
so that we need the response of the stationary distribution $\rho_x\rightarrow \rho_{x+\id x}$ under a change in probe position at $x=0$.  That can be obtained from the linear response theory around steady nonequilibrium as in ref.~\cite{baiesi_jsp_2009}.  The resulting response formula has first a traditional (Kubo-like) entropic part which reproduces the equilibrium form \eqref{stiff-eq} and the second contribution is frenetic and depends on more kinetic details. To be more specific we consider identical medium particles undergoing the overdamped diffusion \eqref{diff} which we write here with $\chi=1$, 
\begin{equation}\label{diff1}
\dot y_t = F(y_t) - \nabla U_x(y_t) + (2T)^{1/2} \xi_t\,,\qquad
\nabla \cdot F = 0
\end{equation}
We  take the potential and driving field
\begin{equation}\label{gen}
U_x(y) = V(|y|) + U_I(|y-x|),\qquad F(y) = \ve |y| \omega(|y|)\,\hat e_\varphi
\end{equation}
for given radial force profile $\omega(r)$.
The backward generator of that driven diffusion equals
\begin{equation}\label{gendr}
L_x = (F - \nabla U_x) \cdot \nabla + T \De = L_x^\text{eq} + \Om
\end{equation}
with, using polar coordinates, $ \Om = F \cdot \nabla = \ve \omega(r)\,\frac{\partial}{\partial \varphi}$.
We easily check the orthogonality relation
\begin{equation}\label{orthogonality}
F \cdot \nabla U_0 = 0
\end{equation}
under which we derive in the Appendix the response formula
\begin{equation}\label{pertu}
\be^{-1} \partial_j \log\rho_0 =
-\partial_j U_0 +\langle \partial_j U_0 \rangle^0 -
\left(L_0^\dagger\right)^{-1} \Om\,(\partial_j U_0)
\end{equation}
for $L_0^\dagger = L_0 - \Omega$ the adjoint of $L_0$ under $\rho_0$ (see Appendix).
Note that the driving $\ve$ does not at all have to be small as we have not been doing perturbation in $F$ but in $x$.\\

By substituting~\eqref{pertu} into \eqref{int}, and with the covariance notation
$\langle u; v \rangle = \langle u v \rangle - \langle u \rangle \langle v \rangle$, we have

\begin{eqnarray}
 M_{jk} &=&\langle \partial_j \partial_k U_0 \rangle^0 - \be
\langle \partial_j U_0;\,\partial_k U_0 \rangle^0 \cr
 & & - \be \Bigl\langle \Om\,(\partial_j U_0)\, \frac{1}{L_0} \partial_k U_0 \Bigr\rangle^0 \label{stiff-ort}
\end{eqnarray}

Equation~\eqref{stiff-ort} is a `general' formula for the stiffness
under the orthogonality condition~\eqref{orthogonality} . The second line of \eqref{stiff-ort} constitutes the frenetic contribution and in general cannot be interpreted  in terms of an effective temperature where we would modify the second term via $\beta \rightarrow \beta_\text{eff}$.
As we will indicate next, that formula can be rewritten and made useful for numerical exploration, and secondly, allows for a direct mathematical proof of the positivity of the Lamb shift under some further dynamical condition.\\
We start with a rewriting.
By rotation symmetry the most general form of the stiffness matrix \eqref{sti} for the probe around $x=0$ is
\begin{equation}\label{ls}
M = \left(
      \begin{array}{cc}
        m & -a \\
        a & m \\
      \end{array}
    \right)
\end{equation}
At equilibrium $a^\text{eq} = 0$, and
\begin{equation}\label{eq-simple}
m^\text{eq} =
\frac{\pi\be}{{\cal Z}_0} \int_0^\infty V' U_I'\,e^{-\be(V + U_I)}\, r \id r\,,\qquad
' = \frac{\partial}{\partial r}
\end{equation}
When $\varepsilon>0$ and for angular rotation $\omega(r)$ there is a Lamb shift $m = m^\text{eq}  + \Delta m$, where $\Delta m$ can be obtained from formula \eqref{stiff-ort}.  We show in the Appendix that it can be expressed as
\begin{equation}\label{lamb-comp-om}
\De m = \frac{1}{2}\ve\be\, \text{Im} \bigl\langle U'_I\, \om \Psi \bigr\rangle^0
\end{equation}
where $\Psi(r)$ solves the differential equation
\begin{equation}\label{ode}
-U_0'\Psi' + \frac{T}{r}(r\Psi')' - \frac{T}{r^2}\Psi + i\varepsilon\omega(r) \Psi = -U_I'
\end{equation}
\begin{figure}[h]
\centering
\includegraphics[width=6 cm]{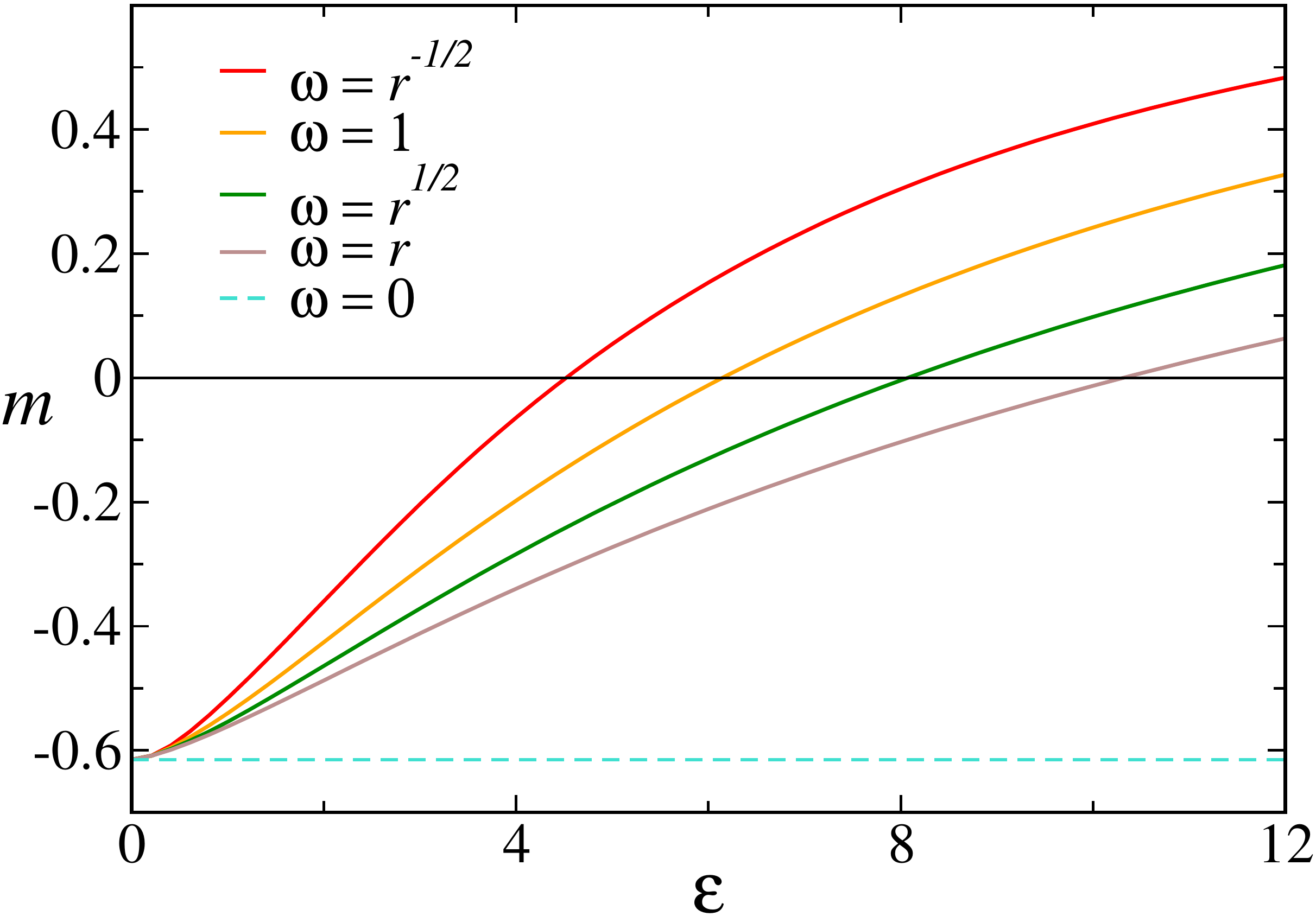}
\caption{Stiffness for various rotation profiles as function of the driving
  $\varepsilon$. Parameters: $T=1$, $\lambda=1$, $k_0=3/4$, $\sigma=1/2$, $\sigma_0=1$, and $R=5$.}
\label{stis}
\end{figure}
which allows direct numerical evaluation.  For example, the stiffness is plotted in
 Fig.~\ref{stis} for various rotation profiles and for
the choice of potentials
\begin{equation}
V(r) = \left\{
\begin{array}{l l}
k_0 e^{-\frac{r^2}{2 \sigma_0^2}} \ &\textrm{for } 0\leq r \leq R\\
0 &\textrm{for } r > R
\end{array}\right.
\end{equation}
\begin{equation}
U_I(r) = -\lambda e^{-\frac{r^2}{2 \sigma^2}}
\end{equation}
The differential equation \eqref{ode} is solved with the routine \texttt{NDSolve} of
Mathematica. The density of the medium is fixed at the boundary of the system as
$\bar\rho(R)=1$.
Note that the Lamb shift is always positive and that there are no dramatic
differences between the types of rotation, including the case
$\omega(r) = 1/\sqrt{r}$.

Secondly, the formula \eqref{stiff-ort} allows a mathematical proof of the positivity of the
nonequilibrium Lamb shift for $\omega(r)=1$.  In that case, $\Omega$ simplifies to $\tilde \Omega =\ve \partial_ \varphi$ and the system enjoys the invariance (see \eqref{proa})
\begin{equation}\label{assumption}
[L_0^\text{eq},\,\tilde \Om] = [L_0,\,\tilde \Om] = 0
\end{equation}
Then $\tilde \Om^\dagger = -\tilde \Om$ (see Appendix) generates a one-parameter symmetry of the equilibrium dynamics for $x=0$ and the symmetric part of $\tilde \Om L_0^{-1}$ is a positive operator as is obvious from rewriting it as
\begin{eqnarray*}
&&\frac{1}{2} \left[ \frac{\tilde \Om}{L_0} + \left( \frac{\tilde \Om}{L_0} \right)^\dagger \right]
= \frac{1}{2} \left[ \frac{\tilde \Om}{L_0} - \frac{\tilde \Om}{L_0^\dagger} \right]
\nonumber\\&&= -\frac{\tilde \Om^2}{L_0 L_0^\dagger} =
\left( \frac{\tilde \Om}{L_0} \right)\, \left( \frac{\tilde \Om}{L_0} \right)^\dagger \geq 0
\end{eqnarray*}
Substituting into~\eqref{stiff-ort}, the symmetric part of the stiffness matrix,
$M_{jk}^{(s)} = [ M_{jk} + M_{kj} ]/2$, obtains the form
\begin{equation}\label{stiff-com}
M_{jk}^{(s)} = M_{jk}^\text{eq} + \be\, \Bigl\langle
\frac{\tilde \Om}{L_0} (\partial_j U_0);
\frac{\tilde \Om}{L_0} (\partial_k U_0) \Bigr\rangle^0
\end{equation}
Or, its nonequilibrium `Lamb shift'  is a positive matrix, symmetric with respect to the driving reversal. The condition \eqref{assumption} is a general sufficient condition leading to \eqref{stiff-com} and to improved stability for dynamics like \eqref{diff}.  Observe that the shift can be interpreted in terms of an excess work because
\begin{equation*}\label{work}
-\tilde \Om\,(\partial_j U_0) =  -F \cdot \nabla (\partial_j U_0) = \partial_j w_0
\end{equation*}
is the gradient at $x=0$ of the mean instantaneous power $w_x = F \cdot (F - \nabla U_x)$ of the driving force.  We see that \eqref{stiff-com} takes the covariance of  the time integrals
\begin{equation*}\label{var}
H_j(y) = \int_0^{+\infty}\id t\, \langle \partial_j w_0(y_t)| y_0=y\rangle^{0} = \frac{\tilde \Om}{L_0} (\partial_j U_0)(y)
\end{equation*}
Alternatively, in \eqref{lamb-comprig} we give the analogue of \eqref{lamb-comp-om}.\\
Note that the forcing $F(y) = \varepsilon\, r\,\hat e_\varphi$ makes a purely rotational field in the sense that its Liouvillian
$\tilde \Omega = \varepsilon \frac{\partial}{\partial\varphi}$
generates rotations around the origin which obviously leave the potential $U_0(y) = U(x=0,y)$ invariant, 
does not imply that the medium satisfies Gibbs rotational ensemble; there is no imposed angular momentum or rigid rotation of a container.

A linear example consists of rotation-symmetric quadratic potentials
$V(r) = \kappa r^2/2,\quad U_I(r) = \la r^2/2\quad
(\kappa + \la > 0)$
for which the equilibrium stiffness~\eqref{eq-simple} is
$m^\text{eq} =
\la \kappa/(\kappa + \la)$.
Equation~\eqref{ode} for  $\omega= 1$  has
the solution $\Psi(r) = \la \,r/(\kappa + \la - i\ve)$
so that the Lamb shift becomes
\begin{equation*}
\De m = \frac{\ve^2 \la^2}{(\kappa + \la)[(\kappa + \la)^2 + \ve^2]}
\end{equation*}
in accord with the results  in ref.~\cite{basu_njp_2015}.
For $0 > \kappa > -\la$ there exists the threshold
$\ve^* = \sqrt{-\kappa(\kappa + \la)}$ such that
$m < 0$ (instability) for $|\ve| < \ve^*$ whereas
$m > 0$ (stability) for $|\ve| > \ve^*$.  We could have thought that linearizing our model would also yield the same stabilization behavior.  However, linearization does not yield the correct statistical force outside equilibrium.  Even for local interactions $U_I$ we can expect a rather strong dependence in the Lamb shift on the medium density far away from the origin.  The reason is that $L_0^{-1}$, just like the Green function of the Laplacian, generally has logarithmic (in two dimensions) or algebraic (in three dimensions) asymptotics.  To make that point clear we give in Fig.~\ref{nonl-t}~(a) the dependence of the
stiffness on changes in the self-potential
\begin{equation}
\label{U_b}
V(r) =  k_0 e^{- \frac{r^2}{2 \sigma_0^2}} + k_w e^{r - \sigma_w} + k_b e^{-\frac{(r - r_b)^2}{2\sigma_b^2}}
\end{equation}
for rotation force $\omega(r)=1$ and for interaction potential
\begin{equation}\label{intera}
U_I(x, y) = - \lambda \left[ 1 - \frac{(x-y)^2}{\sigma^2}  \right]^2
\end{equation}
with a cut-off at $|x-y|=\sigma$.
The change in stiffness is related to a feature of the potential that is located at a radius
of $r_b=4$; the nonlocal dependence of the Lamb shift on far-away features implies that the linearization of the medium dynamics does not produce the correct Lamb shift.

To obtain the stiffness of the probe for nonlinear media, we need direct numerical
simulations of Eq.~\eqref{diff} with a fixed $x$ (with the stochastic Runge-Kutta
algorithm~\cite{branka_heyes_1998}). The force on the probe is obtained as the average over
the stationary regime for a single bath particle.
\begin{figure}[h]
\centering
\includegraphics[width=8.8 cm]{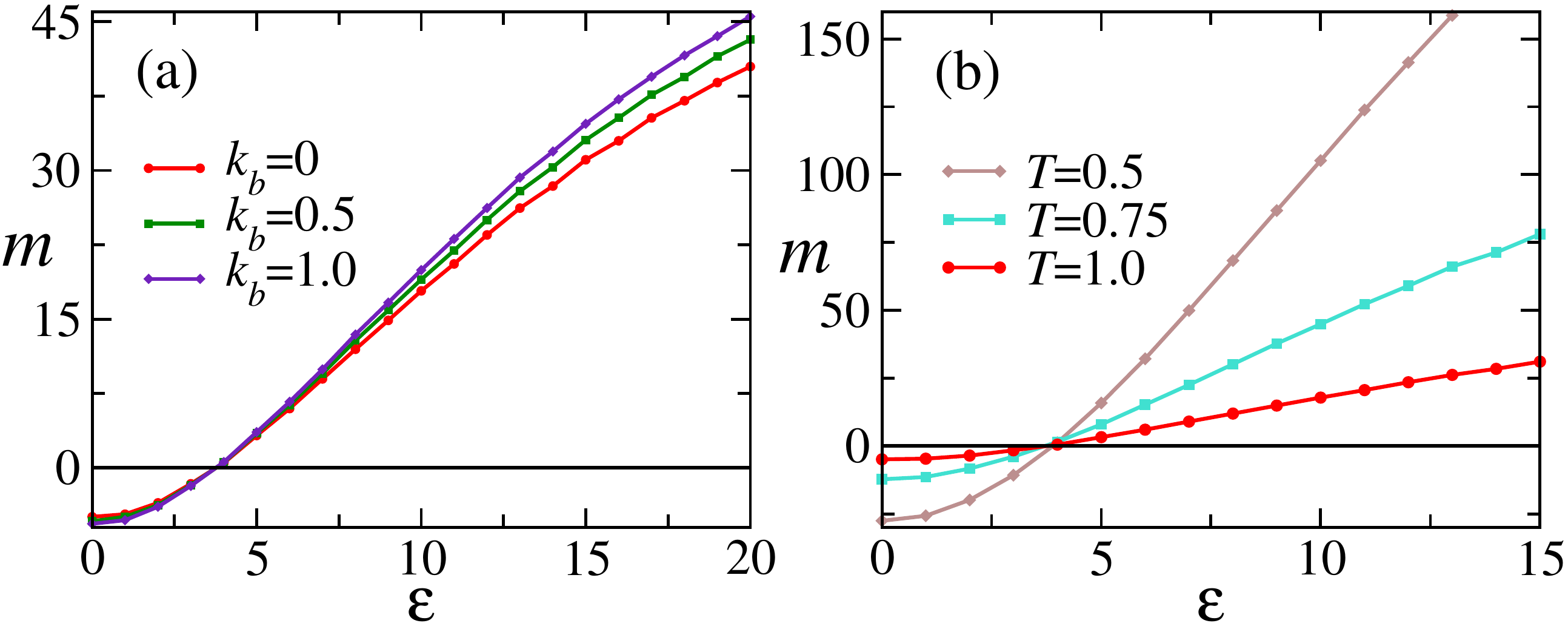}
\caption{(a) Stiffness of the statistical force at the origin for $\omega(r)=1$
  with driving $\varepsilon$. We note the long-range dependence on $k_b$ in the
  self-potential $V$ for large enough $\ve$.  We took $r_b=4, \sigma_b=0.3$.
  (b) Stiffness of the statistical force at the origin for $\omega(r)=1$ with
  driving $\varepsilon$ for various temperatures. Choice of potentials
  is \eqref{U_b} with $k_b=0, k_0=1$ and \eqref{intera} for the interaction.
  }
\label{nonl-t}
\end{figure}
As a further example we give the temperature dependence on the stiffness in Fig.~\ref{nonl-t}~(b).  We
see that the dependence on temperature $T$ is to have greater stability for larger $\ve$ when
$T$ is smaller, but the threshold value varies little with $T$.

\section{Beyond the quasi-static limit}

We return to the coupled system of equations~\eqref{diff}-\eqref{probedyn}.
The medium particles are confined in a disk by a kind of Mexican hat potential \eqref{U_b}
of outer radius 
$\sigma_w$ and with an origin 
of size $\sigma_0$.  As we are interested in studying the stability of the origin,
the probe is also confined to avoid trajectories in which the probe exits the bath region
with a self-potential
\begin{equation*}
\label{U_p}
V_p(x) = e^{|x| - (\sigma_w + 1)}
\end{equation*}
The interaction potential remains the attractive quartic potential of \eqref{intera}.
The driving is via rotation field $F(y) =  \ve |y|\,\hat e_\varphi$.

\begin{figure}[th]
\centering
\includegraphics[width=8.8cm]{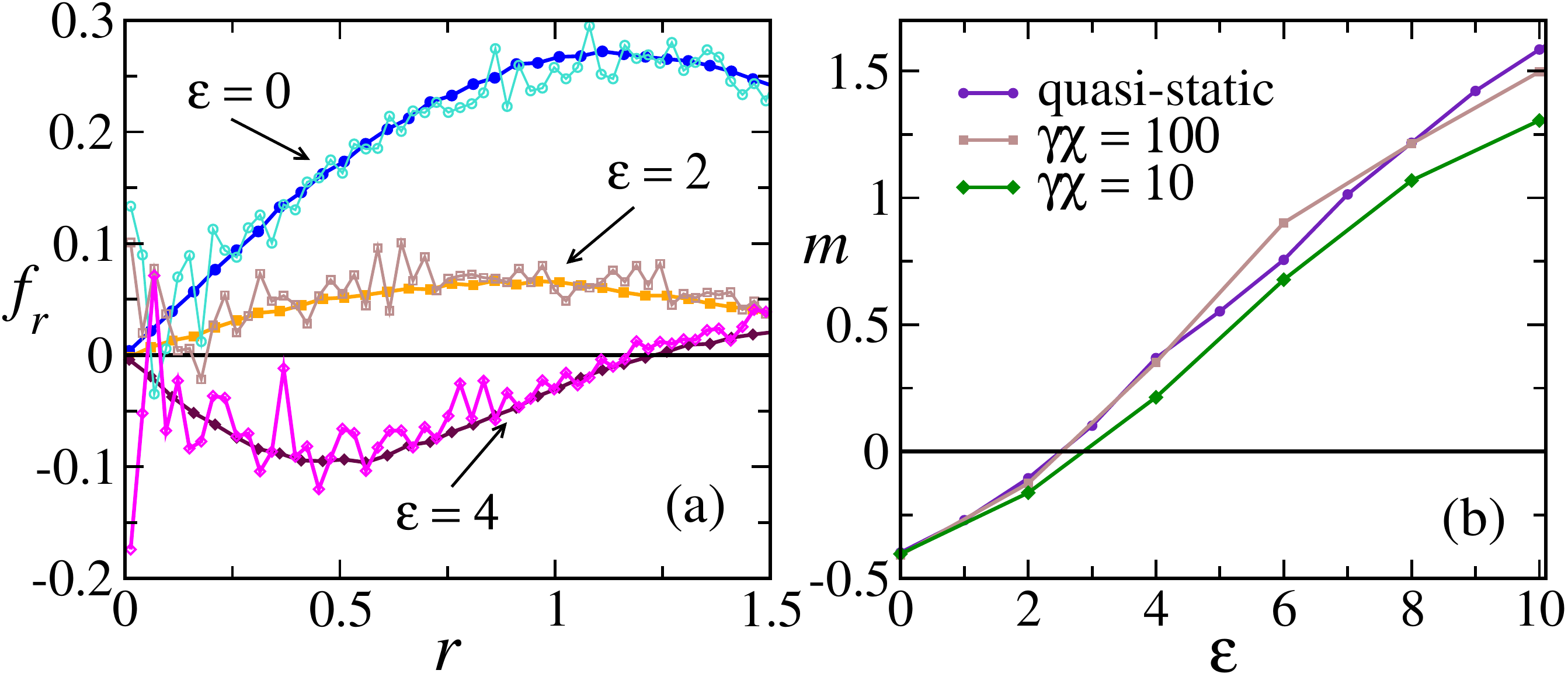}
\caption{(a) The radial statistical force on the quasi-static probe under
  nonequilibrium. Results for quasi-static simulations and full simulations (with
  $\gamma\chi=100$) are superimposed, the latter displaying stronger fluctuations.
  The variation of $\varepsilon$ allows us to see the transition from unstable to
  stable for the fixed point at the origin.
  (b) The stiffness as a function of $\varepsilon$ for both quasi-static and full
  ($\gamma\chi=100$ and $10$) are superimposed.}
\label{stat-force-stiffness}
\end{figure}

We now turn to situations where $\gamma\chi$ is finite with full simulations of
Eqs.~\eqref{diff}-\eqref{probedyn} with the stochastic Runge-Kutta
algorithm~\cite{branka_heyes_1998}. The parameters are given in table~\ref{tab:params}.
A direct comparison between quasi-static and full simulations, for the radial force on the
probe, is shown in Fig.~\ref{stat-force-stiffness}~(a) where good agreement is
found. The results for the full simulations show more fluctuations, related to the
sampling when the probe is moving, with lower radii being most affected.

\begin{table}[h]
\centering
\caption{Parameters for the quasi-static~\eqref{diff} and full \eqref{diff}-\eqref{probedyn}
  numerical simulations. The time step is $10^{-3}$ for the quasi-static simulations and
  $2\ 10^{-3}$ for the full simulations. The parameters in the lower table are used except
  where explicitly stated.}
\begin{tabular}{l l l l}
 & $\gamma$ & $\ve$ & $\lambda$\\
 \hline
 quasi-static - Fig.~\ref{nonl-t} & N/A & 0 to 20 & 5\\
 quasi-static - Fig.~\ref{stat-force-stiffness} & N/A & 0 to 10 & 2\\
 full & 100 & 0 to 10 & 2\\
 full (higher mobility) & 10 & 0 to 10 & 2
\end{tabular}
\vskip 1em
\begin{tabular}{l| l l l l l l l l}
Parameter & $T$ & $\chi$ & $\sigma_w$ & $k_w$ & $\sigma_0$ & $k_0$ & $\sigma$\\
\hline
Value & 1 & 1 & 6 & 1  & 1 & 1/2  & 1\\
\end{tabular}
\label{tab:params}
\end{table}

The stiffness dependence, shown in Fig.~\ref{stat-force-stiffness}~(b) confirms the
agreement with our quasi-static results. Increasing the probe mobility, we can observe that
the behaviour of the stiffness changes and deviates from the quasi-static result, for
increased values of the driving $\varepsilon$. In equilibrium, i.e., for $\varepsilon=0$, we
expect no deviation at all, as observed.

As the probe now moves around, we get access also to
the radial density. In Fig.~\ref{probe_bath_rho}~(a) we observe a dip for
$\varepsilon=0$ that is typical to an unstable fixed point. Increasing $\varepsilon$ leads
to a transformation of this dip into a flat density (for near zero stiffness) and then a
local excess of density (for a stiff origin). 
There is no concurrent change in the local radial density of bath particles $\rho(r)$;
 $\rho(r)$ is displayed in Fig.~\ref{probe_bath_rho}~(b), the
nonequilibrium driving does not change its shape.
\begin{figure}[h]
\centering
\includegraphics[width=8.8 cm]{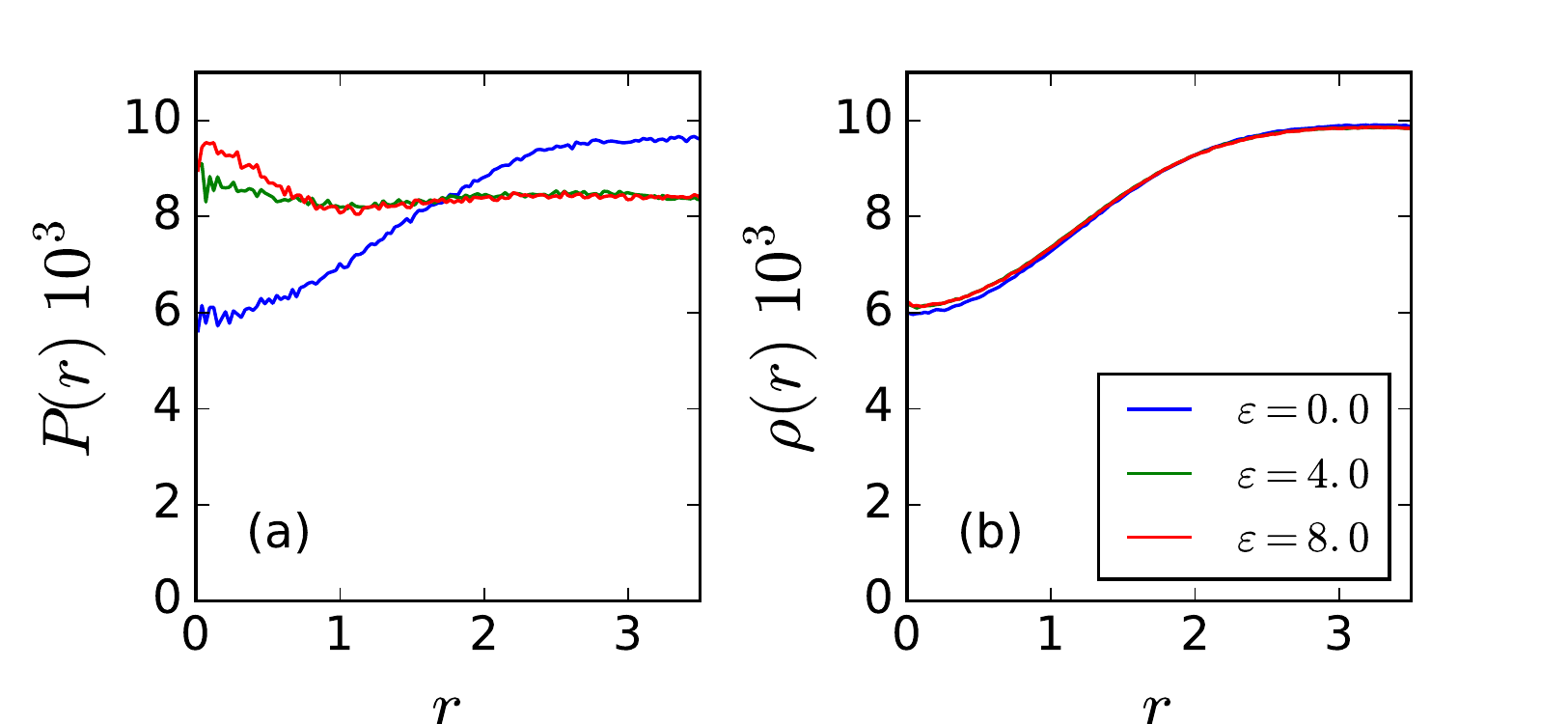}
\caption{(a) The radial distribution $P(r)$ of the probe for full simulations of
  Eqs.~\eqref{diff}-\eqref{probedyn}. (b) The radial distribution $\rho(r)$ of the medium
  for the same simulation set.}
\label{probe_bath_rho}
\end{figure}

\section{Conclusion}

In a nonequilibrium environment different shapes, phases or configurations can become more
stable than in equilibrium.  A systematic treatment uses response theory to check the linear stability around fixed points.
That has been illustrated here, both mathematically and via numerical simulation, to evaluate the nonequilibrium Lamb shift and stiffness of a slow probe in short range interaction with driven medium particles.
There is a simple sufficient
condition for increased stability in terms of the invariance of the equilibrium dynamics under the driving
flow, but our examples showed great robustness of that result beyond the quasi-static limit and for all types of differential rotation.
We have not found a simple heuristics explaining those results; approaches via effective temperature or radial density-profile changes in the driven medium do not appear to work.  At any rate, the Lamb shift in the effective spring constant is second order in the nonequilibrium driving and thus the effect falls outside equilibrium-like energy-entropy  considerations.

\acknowledgments

P.d.B. is a postdoctoral fellow of the Research Foundation-Flanders (FWO). We thank Alexandre Lazarescu for helpful discussions.

\appendix
\section*{Appendix}
\paragraph{Stability}
 The most general notion of stability requires $\lim_{t \to \infty} x_t = 0$ for all $x_0 = x$. That asymptotic stability is equivalent to the strict positivity of the real parts of all eigenvalues of $M$ \
($\Rightarrow$ strict contractivity of the semigroup $e^{-M t}$).  It suffices in general to have a simple  Lyapunov function. (See, e.g., Theorem~3.6 in ref.~\cite{khalil_nonlinear_systems}.)\\
The simplest candidate for a Lyapunov function is
$\lambda(x) = |x|^2$. If the induced probe dynamics is overdamped,
$\gamma \dot x_t = f(x_t)$ with some friction $\gamma > 0$ and linear approximation $f(x) = -Mx$, then
\begin{equation*}
\gamma\, \frac{\id \lambda(x_t)}{\id t} = -2 x_t \cdot M_s x_t
\end{equation*}
and hence $\lambda(x)$ is (exponentially) Lyapunov with attractor at
$x^* = 0$
if and only if $M_s > 0$, i.e., if all its eigenvalues are strictly positive.
Provided that is verified, then the antisymmetric part
$M_a = (M - M^*)/2$ representing rotational forces does essentially not matter for stability (though it of course enters the phase portrait).
If $M$ is a normal matrix, $[M,M^*] = 0 = [M_s,M_a] $, then the real parts of eigenvalues of $M$ coincide with the eigenvalues of $M_s$. In that case, asymptotic stability is equivalent to $M_s > 0$.

\paragraph{Proof of response formula \eqref{pertu}}
The case $x=0$ is a particularly convenient reference for perturbation expansions. The point is that $\rho_0$ equals the equilibrium distribution $\rho_0\propto \exp[-\beta U_0(r)]$ for all $\varepsilon$ because we have assumed that the self-potential is purely radial and therefore $\rho_0$ is also invariant for $\Omega= \varepsilon \omega(r)\,\partial_\varphi$.\\ 
Differentiate $\langle L_x u \rangle^x = 0$ at $x=0$ and use the simplified notation
$\partial_j = \partial / \partial x_j$ and
$\partial_j u_0 = (\partial u_x / \partial x_j)|_{x=0}$: for all functions $u$,
\begin{equation*}
\begin{split}
0 &= \langle (\partial_j \log\rho_0\, L_0 + \partial_j L_0)\,u \rangle^0
\\
&= \langle u L_0^\dagger \partial_j \log\rho_0 - \nabla \partial_j U_0 \cdot \nabla u \rangle^0
\\
&= \langle u\,\{ L_0^\dagger \partial_j \log\rho_0 - \be \nabla U_0 \cdot \nabla \partial_j U_0 + \De \partial_j U_0 \} \rangle^0
\\
&= \langle u\,\{ L_0^\dagger \partial_j \log\rho_0 + \be L_0^\text{eq} \partial_j U_0 \} \rangle^0
\end{split}
\end{equation*}
That yields the equation $L_0^\dagger \partial_j \log\rho_0 = -\be L_0^\text{eq} \partial_j U_0$
or, equivalently,
\begin{equation*}
L_0^\dagger(\partial_j \log\rho_0 + \be \partial_j U_0) =
-\be \Om\, \partial_j U_0
\end{equation*}
the solution of which is \eqref{pertu}.  That immediately gives rise to formula \eqref{stiff-ort}.  To go from there to \eqref{stiff-com} for the case of $\omega(r)=1$, we need the commutativity~\eqref{assumption}, $[L_0, \tilde \Om] = 0$ with $\tilde \Om = \varepsilon \partial_\varphi$ and
\begin{equation}\label{proa}
L_0 = \ve \frac{\partial}{\partial\varphi} -
U_0' \frac{\partial}{\partial r}
+ T\,\Bigl[ \frac{1}{r} \frac{\partial}{\partial r}
\Bigl( r \frac{\partial}{\partial r} \Bigr) +
\frac{1}{r^2} \frac{\partial^2}{\partial\varphi^2} \Bigr]
\end{equation}
In general, for $\Om = \omega(r)\tilde \Om$, $ [L_0, \Om] \ne 0 $ and only the orthogonality
$F \cdot \nabla U_0 = 0$ remains verified.

To show that $L_0^\dagger = L_0^\text{eq} - \Om$ we note, for all $\om(r)$ and arbitrary functions $u$ and $v,$
\begin{eqnarray}
 \langle u\, \Om v \rangle^0 &=&  \int \rho_0^\text{eq}\, u \nabla \cdot (Fv) =  \ve \int \rho_0^\text{eq}~ u \,  \om \, \partial_\varphi v \cr 
 &=& - \ve \int \rho_0^\text{eq}\, v \, \om\, \partial_\varphi u = -\langle v \, \Om u \rangle^0 \nonumber
\end{eqnarray}
i.e., $\Om$ is an antisymmetric operator,
$\Om^\dagger = -\Om$, while by detailed balance
$L_0^\text{eq}$ is symmetric. This means that the (driven) adjoint dynamics for
$x=0$ has the generator
$L_0^\dagger = L_0^\text{eq} - \Om$ which differs from $L_0$ only by the driving reversal.
Therefore the assumption~\eqref{assumption} ensures the normality property,
$[L_0,\,L_0^\dagger] = 0$.

\paragraph{Proof of  \eqref{lamb-comp-om}--\eqref{ode}}

To determine the Lamb shift $\De m$ in \eqref{ls} with respect to equilibrium we first note that
formula~\eqref{stiff-ort} can be written as
\begin{equation*}
\De M_{jk} = 
\Bigl\langle (\om \nabla_j U_I)\,
\frac{1}{L_0^\text{eq} + \om\tilde\Om} \tilde\Om(\nabla_k U_I) \Bigr\rangle^0
\end{equation*}
For the rightmost vector we need, in Cartesian coordinates,
\begin{equation}
\frac{\partial}{\partial \varphi}\,(\nabla U_I) =
U_I'(r)\,\left(-\sin\varphi,\cos\varphi\right)
\end{equation}
and we want to find $h(y) = (h_1(y),h_2(y))$ with $\langle h\rangle^0=0$ so that $(L_0^\text{eq} + \om\tilde\Om)h =- \ve U_I'\,\left(\sin\varphi, -\cos\varphi\right)$.  Going to complex notation, we write $h(y) = \Psi(r) e^{i\varphi}$  and note that for any  $\Phi = \Phi(r)$,
\begin{equation*}\label{lambda}
L_0^\text{eq} (\Phi e^{i\varphi}) = \bigl[
-U_0' \Phi' + \frac{T}{r}(r \Phi')' - \frac{T}{r^2}\Phi \bigl]\,e^{i\varphi}
=: (\La \Phi)\,e^{i\varphi}
\end{equation*}
As a consequence, 
\begin{equation*}
\frac{1}{L_0} \tilde\Om(\nabla U_I) =
\ve\left(\text{Im},-\text{Re}\right)(\Psi e^{i\varphi})
\end{equation*}
where  $\Psi, \langle |\Psi|^2 \rangle^0 < +\infty$  solves
\begin{equation}\label{ode-om}
(\La + i \ve \om)\Psi = -U'_I
\end{equation}
Combining with
$\nabla U_I = \left(\text{Re},\text{Im}\right)(U'_I e^{i\varphi})$ it finally yields \eqref{lamb-comp-om}.\\
Since
$\text{Im}~ \Psi = O(\ve)$, the Lamb shift is $O(\ve^2)$ as expected due to the symmetry $\ve \leftrightarrow -\ve$.\\
In the case where $\omega(r)=1$ we can take $\tilde \Psi$ solving
\begin{equation*}\label{ode1}
(\La + i\ve) \tilde \Psi = -U'_I
\end{equation*}
and the Lamb shift obtains the simplified expression
\begin{equation}\label{lamb-comprig}
\De m = \frac{1}{2} \ve^2 \be\,\bigl\langle |\tilde \Psi|^2  \bigr\rangle^0 
\end{equation}
giving an alternative to \eqref{stiff-com}.

\bibliographystyle{apsrev4-1}
\bibliography{qs_probe}

\end{document}